\documentclass[conference]{IEEEtran}
\usepackage{cite}
\usepackage{graphicx}
\usepackage{mdwtab}
\usepackage{amsmath}
\usepackage{mathrsfs}
\usepackage{amsfonts}
\usepackage{multicol}
\usepackage{algorithm}
\usepackage{algorithmic}
\usepackage{amssymb}
\usepackage{amsthm}
\usepackage{epsfig}
\usepackage{epstopdf}
\usepackage[utf8]{inputenc}
\usepackage[english]{babel}
\usepackage[pages=some,placement=top]{background}

\renewcommand{\qed}{$\blacksquare$}
\begin{document}

\backgroundsetup{contents=This paper has been accepted for presentation and publication in IEEE Globecom 2018 NGNI.,color=black!100,scale=1.25,opacity=0.7,position={6.75,1.75}}
\BgThispage

\title{Shared Spectrum for Mobile-Cell's \\Backhaul and Access Link}

\author{\IEEEauthorblockN{Shan Jaffry, Syed Faraz Hasan and Xiang Gui}
\IEEEauthorblockA{School of Engineering and Advanced Technology,\\
Massey University, New Zealand\\
Correspondence: \{S.Jaffry,F.Hasan,X.Gui\}@massey.ac.nz }
}

\maketitle

\begin{abstract}
Offloading cellular hotspot regions to small-cells has been the main theme for the fifth generation of cellular network. One such hotspot is the public transport which carries a large number of cellular users who frequently receive low quality of service (QoS) due to vehicular penetration effect (VPE). Hence installation of mobile-cell (MC) within public transport is seen as a potential enabler to enhance QoS for commuting users. However, unlike fixed cells, MC requires wireless backhaul (BH) connectivity along with in-vehicle Access-Link (AL) communication. These additional wireless links for MC communication will pose an excessive burden on an already scarce frequency spectrum. Hence, in this research, we exploit VPE and line-of-sight (LOS) communication to allow the downlink backhaul (DL-BH) sub-channels to be shared by in-vehicle downlink access-link (DL-AL) transmission. Our analysis and simulations show that using the above-mentioned technique, both links maintain high success probability, especially in regions with low signal to interference ratios.
\end{abstract}

	
\section{Introduction}	

The main theme for fifth generation (5G) of cellular network pivots around co-existence of multiple radio access technologies \cite{mesodiakaki2017energy}, and cells with different sizes \cite{gupta2015survey}. In such a heterogeneous network (HetNet), the macrocell eNB (MeNB) provides wider coverage area, while small-cells cater dense cellular hotspots \cite{cimmino2014role}.

\begin{figure}[t]
	\centering
		\includegraphics[width=\linewidth]{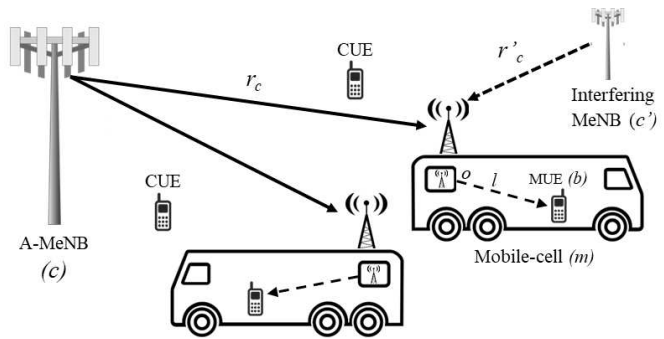}
	\caption{A Mobile-Cell with active backhaul and access link communication.}
	\label{fig:fig_0}
\end{figure}

However, the conventional small-cells could only provide fixed coverage. On the other hand, mobile-platforms like trains, subways, or buses also carry large number of cellular users \cite{Ericsson2015}. The commuters inside these mobile-platforms experience low quality of service (QoS) due to vehicular penetration loss (VPL), which can be as high as 25 dB \cite{tanghe2008evaluation}. Additionally, simultaneous group handovers from public transport generate excessive amount of undesired signaling \cite{lee2018performance,shanPotentials}. Hence researchers aim to unite such users into a single entity as seen by the core-network and call it a mobile-cell (MC). The MC needs to be installed with a bi-antennae transceiver system (see Fig. \ref{fig:fig_0}). The backhaul (BH) antenna will enable wireless connectivity to the core-network. The in-vehicular users will be catered by access-Link (AL) antenna. Researchers in \cite{jaffry2016making, yasuda2015study, jaziri2016offloading, sui2012performance, khan2017outage} have demonstrated that the MC can eliminate VPL, enhance QoS for commuting users, reduce number of handovers, and increase network throughput.

However, as demonstrated in \cite{jaziri2016offloading}, unplanned mobility of MC increases interference to macro-cell layer, decreasing the overall system throughput. 
Chae et al. \cite{chae2012dynamic} proposed to use separate bands for in-vehicle and out-of-vehicle cellular users. This naive approach is less practical since service providers focus on increasing spectral efficiency of their network while maintaining high QoS. Janghser in \cite{jangsher2015resource} combined graph theory with optimization techniques to assign separate power levels and frequency sub-channels to fixed and mobile cells. Same authors in \cite{jangsher2017backhaul} discussed BH spectrum assignment similar to \cite{jangsher2015resource}. Analysis for uplink communication are performed in \cite{khan2017outage} which demonstrated 50\% reduction in outage probability for commuting users. The algorithms presented in \cite{haider2011spectral,jangsher2015resource, jangsher2017backhaul,khan2017outage} assign resources (i.e. sub-channel) to any single MC link. However, since MC bears multiple wireless links (e.g. BH, AL) these approaches  yields additional burden on already scarce spectrum. Hence in this paper, we utilize the inevitable vehicular penetration effect (VPE) to propose resource sharing between downlink backhaul (DL-BH) and downlink access-link (DL-AL) to increase the spectral efficiency without compromising on QoS. We demonstrate through analysis and simulation that with the use of directional antenna (i.e. line-of-sight communication) for DL-AL, along with utilizing VPE, both links can share same resource with high success probability, especially for low signal-to-interference ratio (SIR).  

The main contributions of this research is to study the impact of VPE and LOS communication to enable resource sharing in DL-BH and DL-AL links. An extended version of this research is presented in \cite{shanTransactionSubmitted}.
For the rest of the paper, section \ref{sec_sys_model} and \ref{sec_perf_anal} explain the system model and performance analysis, respectively. Section \ref{sec_results} presents the results and discussions, followed by conclusion in Section \ref{sec_conclusion}.
\section{System Model}
\label{sec_sys_model}

\subsection{Network Model}
In our model, MC communicate with MeNB on BH links. We have considered downlink communication. MeNBs are distributed according to homogeneous poisson point process (PPP) $\Phi_c$ with density $\lambda_c$ (points/m$^2$) in the Euclidean plane. The probability density function (PDF) for the distance between MC $m$ to the nearest base-station is given as \cite{andrews2011tractable}:
\begin{equation}
	f(r) = 2\pi\lambda_{c} r_c \exp(-\lambda_{c}\pi r_c^2).
\end{equation}
where $r_c$ is the distance between nearest base-station and MC $m$. MC $m$ is picked at random from the set of $M$ MCs. 
Since we are considering downlink analysis, the physical locations of cellular users (CUEs) are not taken into account. We define $o$ as the AL-antenna of MC $m$

We have considered deterministic route for MC (e.g. a train,or a subway) \cite{jangsher2015resource} in low SIR region within macrocell layer. 
Following \cite{jangsher2015resource,jangsher2017backhaul} each MC is considered stationary at a given time instance.
MC is linked to its nearest MeNB for BH communication. This setting yields optimal network performance since MC-to-FeNB BH-links will cause large number of handovers due to smaller FeNB coverage. We consider $r_c$ being the distance between the associated MeNB (A-MeNB) and the reference MC. 

We denote the commuting user (MUE) with $b$. Each MC is installed with a bi-antennae system: (i) an external antenna (for BH) called BH-antenna and (ii) a directional antenna mounted under the roof of MC (for AL), called AL-antenna. The position of AL-antenna is such that it enhances the LOS component in access-link communication \cite{rohani2017improving}. The BH-antenna and AL-antenna are internally connected over wired links. The nature of communication over these links (voice calls, data transmission etc.) is out of scope of this study.

\subsection{Channel Model and Spectrum Allocation}

We have two main links under consideration as shown in Fig. \ref{fig:fig_0}. DL-BH link between MeNB and BH-antenna, and  DL-AL between the AL-antenna and MUE $b$.

The radio propagation model consists of large scale and small-scale fading for all links. All the links to the MC BH-antenna (including MeNB-MC and interfering links) follow quasi-static rayleigh fading \cite{jangsher2017backhaul}. The large scale attenuation follows standard pathloss model i.e. $r_t^{-\alpha_i}$, where $r_t$ is the distance between transmitter ($t_x$) and receiver ($t_r$). $\alpha_i$ denotes the non line-of-sight (NLOS) links pathloss exponent.

Due to the nature of AL-antenna, AL follows Rician fading where the Rician K-factor determines the strength of LOS component. The antenna-to-MUE distance is $l$ and $\alpha_o$ is the LOS pathloss exponent. 
$P_c$ and $P_o$ are the transmit powers for MeNB and MC AL-antenna, respectively. Each link is assigned distinct sub-channels. In the cellular layer, sub-channel $\omega$ is allocated by A-MeNB to MC for DL-BH. DL-AL shares the same sub-channel as DL-BH to improve spectral efficiency. 

\subsection{Signal-to-Interference Ratio}

We have considered interference limited environment where noise can be neglected as the signals from interfering transmitters dominate \cite{andrews2011tractable}. We have considered $\Upsilon_1$ being the SIR on sub-channel $\omega$ between DL-BH link between MeNB $c$ and MC $m$. 
\begin{equation}
	\Upsilon_1(\omega,c\to m) = \frac{P_c r_c^{-\alpha_i}h^\omega_{c,m}}{I_C + I_o \varepsilon + I'_o \varepsilon}.
\end{equation}

Note that $\varepsilon$ is the penetration factor such that $0 < \varepsilon \leq 1$. The value of $\varepsilon$ determines the quality of isolation (due to VPE) between BH and AL links. The lower the $\varepsilon$, the better the isolation between two links. $I_C$ denote cumulative interference from interfering MeNBs, i.e. $I_C = \sum\limits_{c' \in \Phi_c \setminus\{c\}} P_c r_c'^{-\alpha_i}h^\omega_{c',m}$ and $I_o = P_o X_d^{-\alpha_i}h_{o,m}^\omega$ is the interference from DL-AL. $h^\omega_{t_x,t_r} \sim \exp(1)$ denotes the exponentially distributed channel gain between any $t_x$ and $t_r$. $r_c'$ denotes the distances to the interfering MeNBs in $\Phi_{c}$. $X_d$ is the distance between the AL-antenna and BH-antenna. $I_o'$ is the total interference from neighboring MC $m'$ ($\forall m'\in M \setminus m$).
Note that the interference from neighboring MC DL-AL is negligible because of very low transmit power. Furthermore, there is lower chance that any nearby MC will be assigned the same set of sub-channels (e.g. near cell edges). Moreover, the main interfering component to the BH-antenna is the spatially closer MC $m$ AL-antenna. Hence, we consider $I'_o \varepsilon \approx 0$ and  $\Upsilon_1$ becomes:
\begin{equation}
\Upsilon_1(\omega,c\to m) = \frac{P_c r_c^{-\alpha_i}h^\omega_{c,m}}{I_C + I_o \varepsilon }.
\end{equation}

We further define $\Upsilon_2$ as the SIR from AL-antenna $o$ to MUE $b$ receiver on sub-channel $\omega$ as:
\begin{equation}
\label{eq_main_SINR2}
\Upsilon_2(\omega,o \to b) = \frac{P_o l^{-\alpha_o}h^\omega_{o,b}}{I_{c} \varepsilon + I'_o \varepsilon^2 },
\end{equation}
where $I_{c}$ is the interference that the MUE experience by cellular transmitters which can be denoted as: $I_{c} = \sum\limits_{c \in \Phi_C} P_c r_c'^{-\alpha_i}h'_{c}$. We have considered that the AL-antenna is mounted under the roof of MC, such that commuting users are in LOS range of the transmitter. 
$I'_o$ is the total interference from the all the neighboring MC transmission to MUE $b$. Note the transmission from neighboring MC DL-AL will experience two VPE signal degradations. Hence, as above, we consider $I'_o \varepsilon^2 \approx 0$. Then Eq. \ref{eq_main_SINR2} becomes:
\begin{equation}
\Upsilon_2(\omega,o \to b) = \frac{P_o l^{-\alpha_o}h^\omega_{o,b}}{I_{c} \varepsilon}.
\end{equation}
  
Since the DL-AL exhibits Rician fading, the channel $h^\omega_{o,b}$ will follow non-central Chi-squared ($\chi^2$) distribution. The PDF for $f_{h^\omega_{o,b}}(h_o)$ can be given according to $\big[$Ch:3, \cite{goldsmith2005wireless}$\big]$ as:
\begin{equation}
\label{eq_Rician}
f_{h^\omega_{o,b}}(h_o) = \frac{K+1}{P_r} e^{ \frac{-KP_r - (K+1)h_o}{P_r}}
 I_0\Big(2\sqrt{\frac{K(K+1)h_o}{P_r}}\Big),
\end{equation}
where  $K$ is the ratio of power for dominant to the scattered component of AL and $I_0(.)$ is the modified Bessel function of the first kind of zeroth order \cite{goldsmith2005wireless}.\\ 

The K-factor determines the strength of LOS component of the signal. For example, K = 0 means the signal follows multipath fading with no dominant LOS component. On the other hand, K = $\infty$ means that a direct LOS component eliminating all scattering waves. The average received power by Rician fading is $P_r = \int\limits_{0}^{\infty} h_o f_{h^\omega_{o,b}}(h_o)dh_o = 2\sigma^2(K+1)$ \cite{goldsmith2005wireless}. If the scattered component of the link is modeled as the Gaussian random variable with the variance $\sigma^2 = 1/2$, then $P_r = K+1$ . Hence, Eq. \ref{eq_Rician} becomes:

\begin{equation}
\label{eq_pdf_g}
f_{h^\omega_{o,b}}(h_o) = \frac{I_0(2\sqrt{Kh_o})}{e^{Kh_o}}. 
\end{equation}

\section{Performance Analysis} 
\label{sec_perf_anal}
This section presents performance analysis for DL-BH and DL-AL for the proposed model. Note that the successful transmission is based on the probability that the SIR achieved at receiver is above a certain threshold level ($\theta$). We have considered this success probability as the main performance metric in this research. It can be mathematically represented as $p = \mathbb{P}[\Upsilon(\omega,t_x \to t_r) > \theta]$, where $\mathbb{P}[.]$ denotes probability of given event. Note that this expression is equal to the complimentary cumulative density function (CCDF) for $\Upsilon_1(\omega,c\to m)$ and $\Upsilon_2(\omega,o \to b)$, respectively. For convenience in notation, we will simply use terms $\Upsilon_1$ and $\Upsilon_2$ for SIR between transmitter $c$ to receiver $m$  and transmitter $o$ to receiver $b$, respectively.

\subsection{Success Probability for Backhaul Link}
\label{sub_sec_p1}
We start with the success probability for DL-BH ($p_1$).
\begin{equation}
\label{main_eq_p_1}
p_1 = \mathbb{E}\  \big[\mathbb{P}[\Upsilon_1 > \theta \ |\ r_c]\big].
\end{equation}
where the expectation $\mathbb{E}[.]$ is with respect to the distance $r_c$ between MC and A-MeNB. For notational convenience we will use $r$ from now on to denote $r_c$. 
\begin{equation}
p_1 = \int_{r>0} \mathbb{P}[\Upsilon_1 > \theta \ |\ r] \ f(r) dr,
\end{equation}
\begin{multline}
\label{eq_p_1_1}
\small p_1 = \int_{r>0}  \mathbb{P} \Big[ h^\omega_{c,m} > \frac{\theta r^{\alpha_i}}{P_c}(I_C + I_o \varepsilon) |\  r\Big] \\\times 2\pi \lambda_c r e^{-\pi \lambda_c r^2} dr,\normalsize
\end{multline}

As mentioned above $h^\omega_{c,m} \sim \exp(1)$, we use the CCDF of $h^\omega_{c,m}$ along with using Laplace transform for random variables $I_o$  and $I_c$. We can re-write the Eq. \ref{eq_p_1_1} as:
\begin{equation}
\label{eq_shan}
\small p_1 = \int_{r>0} e^{-\lambda_c\pi r^2} \mathcal{L}_{I_C}\Big(\frac{\theta r^{\alpha_i}}{P_c}\Big)  \mathcal{L}_{I_o}\Big(\frac{\theta r^{{\alpha_i}}\varepsilon}{P_c}\Big) 2\pi \lambda_c r dr,
\end{equation}

$\mathcal{L}_{I_C} (s)$ is expressed as \cite{andrews2011tractable}:
\begin{equation}
\label{L_IC}
\mathcal{L}_{I_C}(s) = \exp(-\pi r^2\lambda_c \rho(\theta, \alpha)).
\end{equation}
where $\rho(\theta, \alpha_i) = \theta^{2/\alpha_i}\int\limits_{\theta^{-2/\alpha_i}}^{\infty}\frac{1}{1+x^{\alpha_{n}/2}}$.\\

Furthermore, considering $s = \varepsilon \theta (\frac{r}{X_d})^{\alpha_i} \frac{Po}{Pc}$ and using the fact that $\mathcal{L}_h(s) = \frac{1}{1+s}$ \normalsize for $h\sim\exp(1)$, we can find $\mathcal{L}_{I_o}(s)$ as

\begin{equation}
\label{L_IO}
\mathcal{L}_{I_o}(s) = \frac{1}{1 + \varepsilon \theta (r/X_d)^{\alpha_i} P_o/Pc}.
\end{equation} 

Then the success probability ($p_1$) in Eq. \ref{eq_shan}  becomes


\begin{multline}
p_1 = \exp(-\pi r^2\lambda_c \rho(\theta, \alpha)) \times \\\frac{1}{1 + \varepsilon \theta (r/X_d)^{\alpha_i} P_o/Pc} \times 2\pi\lambda_c \exp(-\pi \lambda_c r^2),
\end{multline}

\begin{equation}
= 2\pi \lambda_c \int_{r>0}  \frac{e^{-\pi \lambda_c (1 + \rho(\theta, \alpha_i))r^2}}{1 + \varepsilon \theta (r/X_d)^{\alpha_i} P_o/Pc} r dr,
\end{equation}

\begin{equation}
\label{eq_unsolvable}
p_1 = 2\pi \lambda_c \int_{r>0}  \frac{e^{- \mathcal{Z} r^2}}{1 + \mathcal{Y}r^{\alpha_i}} r dr,
\end{equation}

where $\mathcal{Z} = \pi \lambda_c (1 + \rho(\theta, \alpha_i))$ and $\mathcal{Y} = \varepsilon \theta  P_o/(X_d)^{\alpha_i}Pc$\\

It is difficult to find the closed form expression for Eq. \ref{eq_unsolvable}. However considering $r = \tan\big(\frac{\pi \delta}{2}\big)$, we can rewrite it as

\begin{multline}
\label{eq_unsolvable_2}
p_1 = \pi^2 \lambda_c \int_{0}^{1}  \sec^2\Big(\frac{\pi}{2}\delta\Big) \tan\Big(\frac{\pi}{2}\delta\Big)\times\\ \frac{\exp\Big\{-\mathcal{Z}\tan\big(\frac{\pi}{2}\delta\big)\Big\}}{1 + \mathcal{Y}\tan^{\alpha_i}\big(\frac{\pi}{2}\delta\big)} d\delta.
\end{multline}

For $\alpha_{n} = 4$, $\mathcal{Z} = \pi \lambda_c (1 + \sqrt{\theta}(\frac{\pi}{2} - \tan^{-1}(\frac{1}{\sqrt{\theta}}))$ and \linebreak $\mathcal{Y} > 0$. Eq. \ref{eq_unsolvable_2} can easily be solved using numerical integration methods.


\subsection{Success Probability for Access Link}
\label{sub_sec_AL_prob}
As mentioned in Section \ref{sec_sys_model}, the downlink between AL-antenna and MUE inside MC has strong LOS component due to the use of directional antennas. Hence DL-AL follows Rician distribution. The probability for successful AL transmission can be given as: 
\begin{multline}
\label{eq_p2_final2}
p_2 = \exp{(-\mathcal{X}\theta^{2/\alpha_i})}\sum\limits_{j=0}^{J} \sum\limits_{m=0}^{j} \frac{K^j (-1)^{j-m}}{e^K j!(j-m)!} \\\sum\limits_{q=0}^{Q} \frac{\Gamma(\frac{2q}{\alpha_i} + 1)}{\Gamma(\frac{2q}{\alpha_i} - (j-m) + 1)}. 
\end{multline}
where $\Gamma(.)$ is the Gamma function and $J,Q$ are positive integers such that $\frac{1}{J!}, \frac{1}{Q!} \to 0$

\begin{equation}
\mathcal{X} = \lambda_c\pi (\varepsilon l^{\alpha_o})^{2/\alpha_i} \Big(\frac{P_c}{P_o}\Big)^{2/\alpha_i} \beta(\alpha_i).
\end{equation}
where $\beta(\Delta) = \frac{2\pi/\Delta}{\sin(2\pi/\Delta)}$.\\

See Proof: Appendix \ref{app_second}.

\subsection{Ergodic rate per Sub-channel for DL-BH}

The ergodic rate for BH-link is derived using techniques in \cite{andrews2011tractable} and is presented as Eq. \ref{eq_Rates_3}. Due to page limitations, the proofs are omitted here. Note that $\mathcal{F} = \frac{P_o \varepsilon}{P_c X_d^4}$ in Eq. \ref{eq_Rates_3}.\linebreak

\begin{figure*}[t]
	\hrule
	\begin{equation}
	\label{eq_Rates_3}
	\small T = \int\limits_{g=0}^{g=1} \int\limits_{\sigma=0}^{\sigma=1} \exp\bigg\{-\bigg(\frac{1}{\sigma}-1\bigg)\bigg[ 1 + \sqrt{e^{\frac{1}{g}-1}-1}\bigg(\frac{\pi}{2} - \tan^{-1}\frac{1}{\sqrt{\exp(1/g-1) - 1}}\bigg)\bigg] \bigg\} \times \frac{1}{g^2 \sigma^2 \bigg[1 + \big(\frac{1}{\sigma}-1\big)^2\bigg( \frac{\mathcal{F}\exp(1/g -1) - 1}{\lambda_c^2 \pi^2} \bigg)\bigg]}\ d\sigma\ dg.
	\end{equation}
	\hrule
\end{figure*}

\begin{figure}[t]
	\centering
		\includegraphics[width=\linewidth]{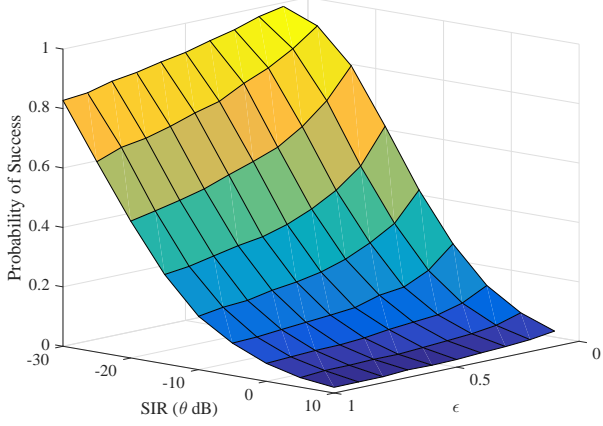}
	\caption{BH-DL success probability vs SIR and $\varepsilon$ \footnotesize($\lambda_c=6\times10^{-6}$)\normalsize}
	\label{fig:fig_3c}
\end{figure}

\begin{figure}[t]
	\centering
		\includegraphics[width=\linewidth]{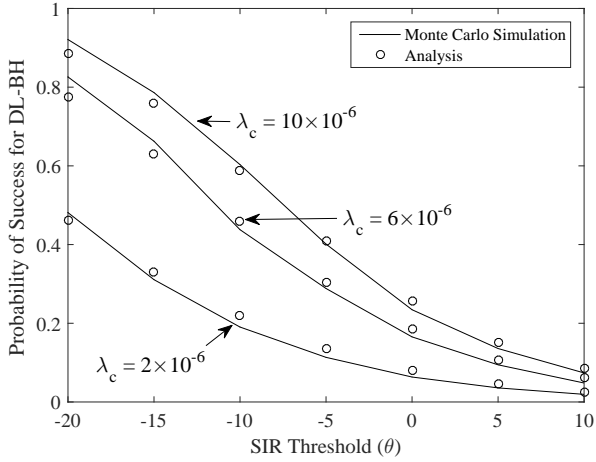}
	\caption{BH success probability with different $\lambda_c$ (points/m$^2$) ($\varepsilon = 0.1$)}
	\label{fig:fig_3d}
\end{figure}

\section{Simulation and results}
\label{sec_results}

In this section, we present the evaluation for the analysis done in previous section using Monte Carlo simulation. The total simulations area is considered to be 40$\times$40 km$^2$. MC and user association to the closest to the MeNB corresponds to Voronoi tessellation for PPP $\Phi_c$. We averaged the results for 10,000 realizations for each simulation and found that the simulation and analysis closely match. General simulation parameters are shown in Table \ref{table_sim_param}.  

\begin{figure}[t]
	\centering
	\includegraphics[width=\linewidth]{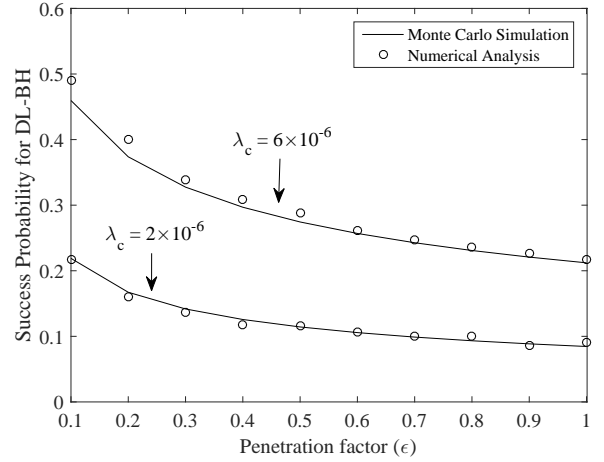}
	\caption{Success probability vs Penetration-factor (SIR = $-10$ dB). }
	\label{fig:fig_3e}
\end{figure}

\begin{figure}[t]
	\centering
		\includegraphics[width=\linewidth]{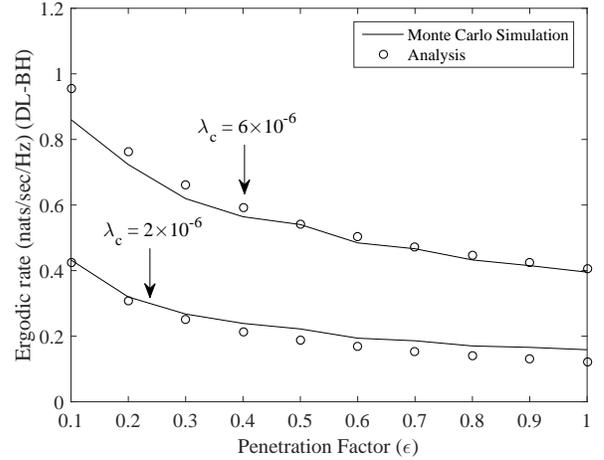}
	\caption{Ergodic rate for backhaul link in nats/sec/hz. }
	\label{fig:fig_3f}
\end{figure}

\begin{table}[t]
	
	\caption{Simulation Parameters} 
	\label{table_sim_param}
	\centering
	
	\begin{tabular}{l l }
		\hline
		Parameters & Numerical value(s) \\
		\hline  			
		Simulation Runs & 10,000\\
		Simulation Area & $40 \times 40$ sq. km\\
		Transmit powers ($P_c, P_o$) & 45 dBm, 3 dBm\\
		Max. AL-antenna $\leftrightarrow$ MUE distance& 8 meters\\
		BH $\leftrightarrow$ AL antennae distance& 5 meters\\
		$\alpha_i$ / $\alpha_o$  & 4/3.5 \\
		$J,Q$ & 70\\
		\hline
		
	\end{tabular}
\end{table}
Fig. \ref{fig:fig_3c} demonstrates the combined effect of penetration factor and SIR threshold on the success probability of BH link. It can be observed that with better isolation between BH and AL (i.e. lower values of $\varepsilon$), greater success probability can be achieved even at higher SIR threshold. This can be achieved with the installation of directional AL-antenna, coupled with better interference cancellation techniques and MC-structural design favoring VPE. Fig \ref{fig:fig_3d} shows that the simulation results closely match the mathematical analysis. The slight difference is due to the use of approximations made for numerical integration in the analysis. It is evident from Fig. \ref{fig:fig_3d} that with higher macrocell density ($\lambda_c$), better BH-link performance can be achieved. This result is also in-line with \cite{yasuda2015study} and intuitively suggests that with the decrement of link-distance between A-MeNB and MC BH-antenna, the success probability increases. Similar results can be observed in Fig. \ref{fig:fig_3e} that increasing $\lambda_c$ improves the performance of  BH, even with highly interfering AL-links. Fig. \ref{fig:fig_3f} demonstrate the affects of VPE and macrocell density on the ergodic rate of BH in nats/sec/hz.

\begin{figure}[t]
	\centering
	\includegraphics[scale=0.4, width=\linewidth]{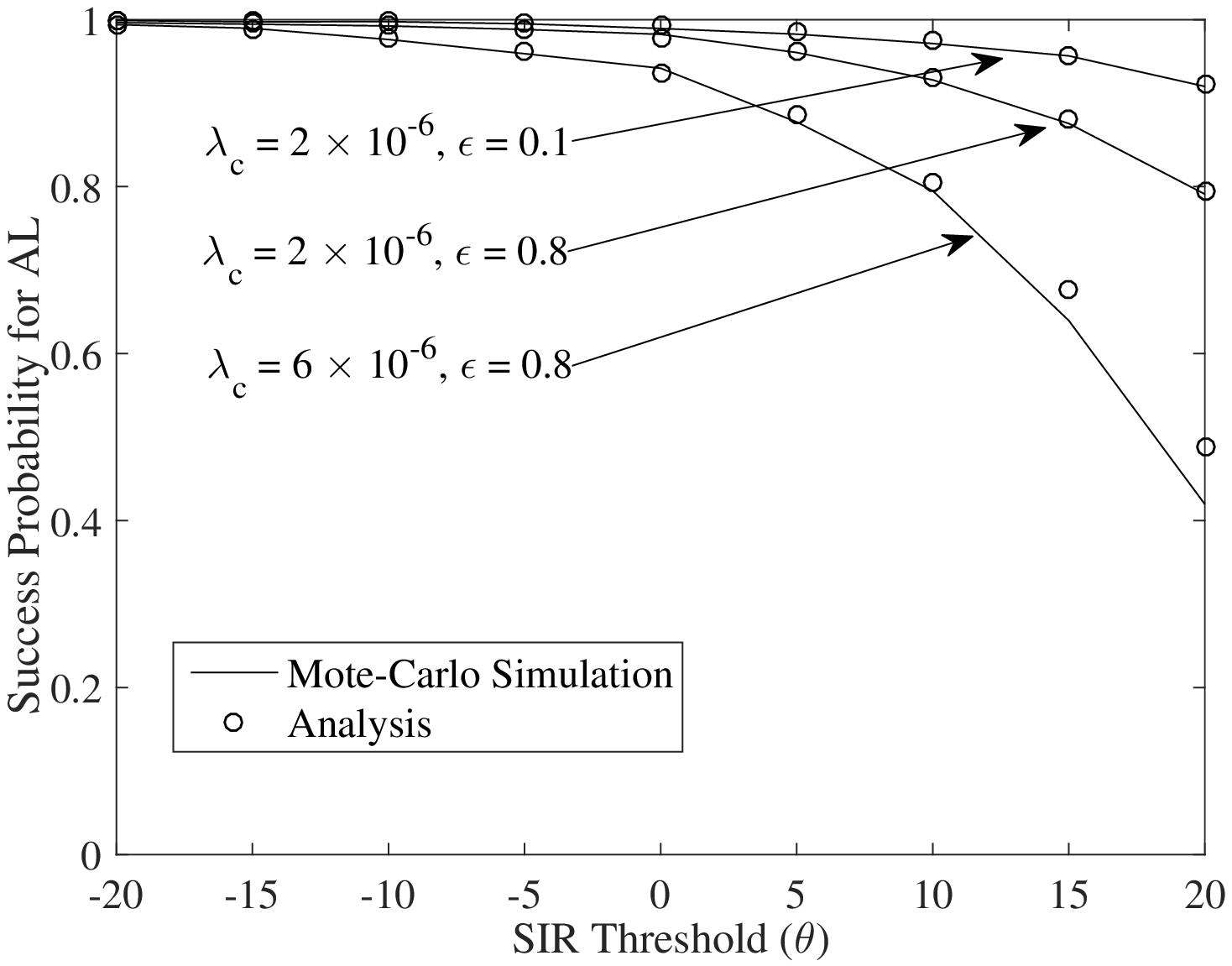}
	\caption{Successful AL transmission ($K$ = 2). }
	\label{fig:fig_4}
\end{figure}

Finally, the performance gains of AL-antenna and MUE are observed in Fig. \ref{fig:fig_4}. As demonstrated in the figure, for a well isolated MC-structure (e.g. $\varepsilon = 0.1$), higher probability of DL-AL transmission is achieved. Due to the LOS component in DL-AL, even higher values of $\varepsilon$ do not result in poor-connectivity. Note however that for a dense macrocell deployment, DL-AL success probability gets low. This is due to that fact that the interfering MeNBs are now spatially closer to the MUE $b$. On the contrary, as demonstrated in Fig. \ref{fig:fig_3d} and Fig. \ref{fig:fig_3e}, the success probability for backhaul link increases with the density of MeNBs. The evaluation of optimal macro-cell deployment is out of scope of this research.

Note in Fig. \ref{fig:fig_4} that Rician K-factor is assumed to be 2 dB, which means LOS component is twice as strong as multi-path components. This is true for in-vehicular users since the directional AL-antenna is mounted under vehicle-roof and hence dominant-signals' effects on communication is more stronger than scattered components. However, the signal strength is reduced to lower the success probability for poorly-isolated MC structural design (e.g. $\varepsilon = 0.8$). 
On the contrary, if we assume a strong LOS then the success probability is very high, even for poorly-isolated scenarios (not shown in the figure). These results also demonstrate that the position of AL-antenna, along with proper seating arrangements inside MC will have significant affects on the performance of MC. Note that in practical scenarios, having very high value of $K$ (e.g. $K\geq10$) is difficult to achieve. 

\section{Conclusion}
\label{sec_conclusion}

In this paper, we have presented a shared DL-BH and DL-AL resource sharing scheme for mobile-cell, exploiting vehicular penetration effect and directional antenna system for downlink access link  communication. Unlike other resource allocation schemes proposed for mobile-cells, we have demonstrated that exploiting VPE, along with using directional antennas for AL, MC DL-BH and DL-AL communication can be performed without using additional resources. While VPE depends upon the material and construction properties of the transport-vehicles, the LOS communication can be enabled using directional antennas inside MC. This paper also gives an idea that construction parameters and antenna positioning inside MC should be considered for future transport vehicles to increase the spectral efficiency for cellular network. Resource sharing for sidehaul links and uplink backhaul and access links are the topic of our on-going research. 

\section{Acknowledgment}
This work has been partially funded through Massey University Conference Grant, 2018.

\appendices

\section{Success Probability for Access link}
\label{app_second}
We start by stating that $p_2$ is the probability of success for MC-to-MUE link which is given as $p_2 = \mathbb{P}(\Upsilon_2 > \theta)$:
\begin{equation}
p_2 = \mathbb{E} \Bigg[1 - \mathbb{P}\Big[h_o \leq \theta \varepsilon I'_{c}\Big] \Bigg].
\end{equation}
where $\mathbb{P}\Big[h_o \leq \theta \varepsilon I'_{c}\Big]$ is the cumulative distribution function (CDF) for the random variable $h_o$ given as $F_{h^\omega_{o,b}}(y)$. Let $I'_{c} = I_{c}/P_o l^{-\alpha_o}$. It is difficult to find the CDF of $h_o$ due to the presence of zeroth order Bessel function in its PDF given in Eq. \ref{eq_pdf_g}. Following \cite{peng2014device}, and after using expansion series provided in [8.447.1 in \cite{gradshteyn2014table}], the PDF is expressed as:
\begin{equation}
f_{h^\omega_{o,b}}(h_o) = \sum\limits^{\infty}_{j=0} \frac{(Kh_o)^j}{e^{(K+h_o)}(j!)^2},
\end{equation}

Based on above equation, the CDF is derived as:
\begin{equation}
F_{h^\omega_{o,b}}(y) = \sum\limits^{\infty}_{j=0} \frac{(K)^je^{-K}}{(j!)^2} \int\limits_{0}^{x}y^je^{-y} dy,
\end{equation}

Then, applying 2.321.2 in \cite{gradshteyn2014table}
$\int x^ne^{ax} = e^{ax}\Big(\sum\limits_{m=0}^{n}\frac{(-1)^m m! \binom{n}{m} }{a^{m+1}}x^{n-m} \Big)$, we get:
\begin{equation}
F_{h^\omega_{o,b}}(y) = \sum\limits_{j=0}^{\infty}\sum\limits_{m=0}^{j} \frac{e^{K}}{K^j j!(j-m)!} e^{-y} y^{j-m} \ ,
\end{equation}

Note that  $y = \theta\varepsilon I'_{c}$, we get $p_2$ as: 
\begin{equation}
\label{eq_ref_1}
p_2 = \sum\limits_{j=0}^{\infty}\sum\limits_{m=0}^{j} \frac{e^{K}}{K^j j!(j-m)!} \int^\infty_0 e^{-y} y^{j-m} f_{I'_{c}}(y) dy\ ,
\end{equation}

Now let $y' = I'_{c}$ and $n = j-m$, Eq. \ref{eq_ref_1} becomes \cite{peng2014device}:
\begin{equation}
\label{eq_p2_intermediate}
p_2 =  \sum\limits_{j=0}^{\infty}\sum\limits_{m=0}^{j} \frac{e^{K}}{K^j j!(j-m)!}(\theta \varepsilon I_{c}')^{n} \int\limits_{0}^{\infty}e^{-\theta y'} y'^n  f_{I'_{c}}(y') dy',
\end{equation}

Using property $\mathcal{L}[t^nf(t)] = (-1)^nF^n(s)$ and $\textit{a} = \theta$ we can simplify Eq. \ref{eq_p2_intermediate} as:
\begin{equation}
p_2 =  \sum\limits_{j=0}^{\infty}\sum\limits_{m=0}^{j} \frac{e^{K}}{K^j j!(j-m)!}(\theta)^n \mathcal{D}(a,n),
\end{equation}
where
\begin{equation}
\label{equal_1}
\mathcal{D}(a,n) = \int\limits_{0}^{\infty}e^{-y'} y'^n  f_{I'_{c}}(y') dy' = (-1)^n D^n \mathcal{L}_{I'_{c}}(a),
\end{equation}

The combined Laplace transform of $I'_{c}$ is derived using methods presented in Section \ref{sub_sec_p1}:
\begin{multline}
\label{eq_ref_2}
\mathcal{L}_{I'_{c}}(\theta) = \exp \Bigg\{-\pi (\theta \varepsilon l^{\alpha_o})^{2/\alpha_i}  \Bigg(\lambda_c\Big(\frac{P_c}{P_o}\Big)^{2/\alpha_i} \Bigg) \beta(\alpha_i) \Bigg\},
\end{multline}
and $D^n(.)$ is the $n_{th}$ derivative of the function and we find $\mathcal{D}(a,n)$ as:
\begin{equation}
 \mathcal{D}(a,n)  = (-1)^n\sum_{q=0}^{\infty}\frac{(-1)^{q} \mathcal{X}^q}{q!} a^{\frac{2q}{\alpha_i} - n}\frac{\Gamma(\frac{2q}{\alpha_i} + 1)}{\Gamma(\frac{2q}{\alpha_i} -n + 1)},
\end{equation}
where $\beta(.)$ and $\mathcal{X}$ are defined in section \ref{sec_perf_anal}.

Finally, $p_2$ becomes:
\begin{multline}
\label{eq_p2_final_0}
p_2 = \sum\limits_{j=0}^{\infty} 	\sum\limits_{m=0}^{j} \frac{K^j (-\theta)^{j-m}}{e^K j!(j-m)!} \sum\limits_{q=0}^{\infty} \frac{(-1)^q \mathcal{X}^q}{q!}\theta^{\frac{2q}{\alpha_i} - (j-m)}\\\frac{\Gamma(\frac{2q}{\alpha_i} + 1)}{\Gamma(\frac{2q}{\alpha_i} - (j-m) + 1)}. 
\end{multline}
\begin{multline}
\label{eq_p2_final}
p_2 = \exp{(-\mathcal{X}\theta^{2/\alpha_i})}\sum\limits_{j=0}^{\infty} 	\sum\limits_{m=0}^{j} \frac{K^j (-1)^{j-m}}{e^K j!(j-m)!} \\\sum\limits_{q=0}^{\infty} \frac{\Gamma(\frac{2q}{\alpha_i} + 1)}{\Gamma(\frac{2q}{\alpha_i} - (j-m) + 1)}. 
\end{multline}

Eq. \ref{eq_p2_final} is still difficult to solve when $j,q\to\infty$. However for large values of $j$ and $q$, $\frac{1}{j!}$ and $\frac{1}{q!}$ will reach zero. Hence we can determine the upper limits for index parameters $j$ and $q$ as $J$ and $Q$, respectively. Note that $J,Q$ must satisfy the condition that $\frac{1}{J!} \to 0$ and $\frac{1}{Q!}\to0$. Hence, we can rewrite the Eq. \ref{eq_p2_final} as Eq. \ref{eq_p2_final2}. 
This completes the proof \qed

\bibliographystyle{IEEEtran}
\bibliography{Globecom2018_references}

\end{document}